\documentclass[a4paper,10pt]{article}
\usepackage{graphicx}
\usepackage{color}

\title{Pacman Percolation and the Glass Transition}

\author{Raffaele Pastore$^{\rm b}$ $^{\ast}$\thanks{$^\ast$Corresponding 
author. Email: pastore@na.infn.it \vspace{6pt}},
Massimo Pica Ciamarra$^{\rm b}$ and
Antonio Coniglio$^{\rm a,b}$
\\\vspace{6pt} $^{\rm a}${\em{University of Naples "Federico II"}}; 
$^{\rm b}${\em{CNR--SPIN, Naples, Italy.}}
\\\vspace{6pt}
}

\def\rp{p}

\def\xip{{\xi_{\rm per}}}

\def\tp{{t_{\rm per}}}

\def\rka{{\rho_{\rm ka}}}
\def\<{\langle}
\def\>{\rangle}
\begin{document}

\maketitle

\begin{abstract}
We investigate via Monte Carlo simulations the kinetically constrained
Kob-Andersen lattice glass model showing that,
contrary to current expectations, the relaxation process and
the dynamical heterogeneities seems to be characterized by different time scales. 
Indeed, we found that the relaxation time is related to a reverse percolation transition, 
whereas the time of maximum heterogeneity
is related to the spatial correlation between particles. 
This investigation leads to a geometrical interpretation of the 
relaxation processes and of the different observed time scales.
\end{abstract}


\section{Introduction}
The main challenge in the field of supercooled liquids is the understanding of
the rapid increase of relaxation time and viscosity as temperature decreases ~\cite{Angell, Arxiv}.
Experimental evidences and theoretical models predict the simultaneous emergence of 
Dynamical Heterogeneities (DHs) \cite{ToninelliE, Ediger}, 
which play the role of critical fluctuation in ordinary critical phenomena.
Indeed, these studies depict DHs as dynamically correlated clusters:
by lowering the temperature, the size of such clusters grows and
their cooperative rearrangement becomes more complicated. 
This results in a rapid increase of typical cluster life-time, 
which is in turn related to the structural relaxation time, at a macroscopic level.
Being defined as the volume integral of a four-point correlation functon, $g_{4}(r,t)$, 
the dynamical susceptibility $\chi_{4}(t)$, estimates the volume of these clusters.
Thus, $\chi_{4}$ is expected to grow as some power of the dynamical correlation length, $\xi_{4}$,
derived by $g_{4}(r,t)$ ; 
in particular $\chi_{4}(t) \propto \xi_{4}(t)^{d}$ 
for compact clusters in $d$ dimensions, whereas a smaller
exponent is expected if clusters have a fractal structure. 
The time $t^{*}$, where $\chi_{4}(t)$ reaches its maximum, $\chi^*_4$,
is also the time where correlations start to decrease,
and thus it is interpreted as an estimate of the typical cluster life-time. 
Then, if DHs and relaxation are strictly related,
$t^{*}\propto \tau$ is expected.
However, there exist systems where $\chi^*_4$
is found to decrease on approaching the transition~\cite{glotzer2000,fierro,drop},
as well as early studies suggesting that the time of maximal correlation between particles displacements
does not scale with the relaxation time $\tau$~\cite{Glotezer_times}.
Accordingly, the relation between the dynamical susceptibility and the relaxation process
remains elusive. Moreover, in order to unveil the precise relation between $\chi_{4}$ and $\xi_{4}$ one
needs to know the explicit form of $g_{4}(r,t)$, which is not an easy experimental task.
 
Here we address this problem via a numerical study of the Kob--Anderson kinetically constrained
lattice gas model~\cite{KA} (KA-model), where it is possible to obtain very accurate data
for the dynamical correlation length.
After shortly reviewing the numerical model (Sec. \ref{model}), we illustrate the relaxation process (Sec. \ref{relaxation})
and the DHs (Sec. \ref{DH}). 
In Sec. \ref{dd} we show that the emergence of DHs is well described by the diffusing defect picture,
which correctly predicts the exponents characterizing DHs.
In Sec. \ref{percolation} we explain the relaxation process in terms of "pacman" percolation, a model originally
introduced to explain degradation of a gel due to the action of enzymes \cite{Abete}. 
This picture furnishes a geometrical interpretation of the relaxation process and of the different observed timescales.
This paper is based on the elaboration of results previously published ~\cite{thesis, PRL2011, Varenna}.

\section{Model and simulation details}
\label{model}
The KA-model~\cite{KA} is a kinetically constrained model~\cite{Ritort},
which consists in a cubic lattice of volume $V = L^3$ containing $N$ particles.
Periodic boundary conditions are imposed along the $x$, $y$ and $z$
directions. The global density $ \rho = N/V $ is the only control parameter and
it plays the role of an inverse temperature.
No interactions between particles are present apart from an hard-core
repulsion, which prevents more than one particle to occupy the same lattice
site: all microscopic configurations where the particles occupy $N$ among
$V$ available sites are allowed, isoenergetic and equiprobable. Thus, the
model is characterized by a trivial energy landscape, where no thermodynamic
transition is available and time translational invariance holds.
The system evolves following dyamical rules based on kinetic constraints: a continous time
stochastic process allows a particle to move in a near empty site if has less than $m = 4$ neighbors,
and if it will also have less than $m = 4$ neighbors after the move. Previous studies
have shown that this model reproduces many aspects of glass forming systems:
the dynamics slows down on increasing the density, suggesting the existence of a transition of
structural arrest at $\rho_{ka} = 0.881$~\cite{KA, FranzMulletParisi, Dawson, Chaudhuri}.
Nevertheless, it has been demonstrated that in the thermodynamic limit the transition of dynamical arrest only occurs
at $\rho = 1$~\cite{Toninelli}.

Our simulations of the KA-model span a set of density values ranging over up $\rho=0.87$.
We have investigated a wide range of system sizes from $L=8$ up to $L=50$.
Data reported in the following concern a system size $L=30$, where we have performed the largest statistics:
for each density values, the results are averaged on at least $10^{2}$ over up $10^{4}$ runs.
The runs at higher density values last up $10^{8}$ Monte Carlo sweeps.

\section{Structural Relaxation}
\label {relaxation}
\begin{figure}[t!]
\begin{center}
\includegraphics*[scale=0.85]{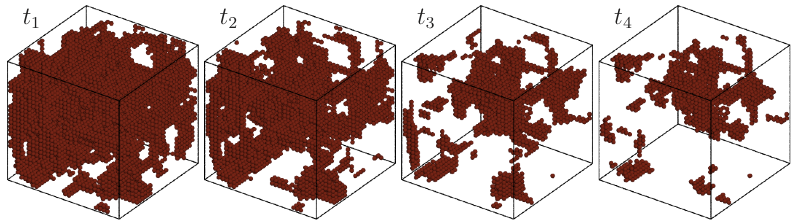}
\end{center}
\caption{\label{fig:plots}
Persistent particles in a numerical simulations of the Kob--Andersen model at $\rho = 0.85$, at times
$t_1 = 3.5~10^5$, $t_2 = 7.5~10^5$, $t_3 = 1.6~10^6$, and $t_4 = 2.1~10^6$. From Ref. \cite{PRL2011}.
}
\end{figure}
Here we investigate the relaxation process focusing on the time evolution
of the density of persistent particles $\rp(t)=\frac{1}V \sum_{i=1}^{V} n_{i}(t)$,
where  $n_{i}(t)$ is known as 'persistence': $n_i(t) = 1(0)$ if site $i$ is (is not)
persistently occupied by a particle in the time interval $[0,t]$ \cite{nota}.
$p(t)$ is related to the high wave vector limit of the intermediate
self scattering function~\cite{Chandler2006} and describes the relaxation of the system in a very direct way: 
initially $p(0)=\rho$, but as time proceeds, particles eventually 
move from their original positions and $p(t)$ decreases.  
This process is represented in Fig. \ref{fig:plots} which shows the 
persistent particles in the system at different times of the same run.

\begin{figure}[t!]
\begin{center}
\includegraphics*[scale=0.5]{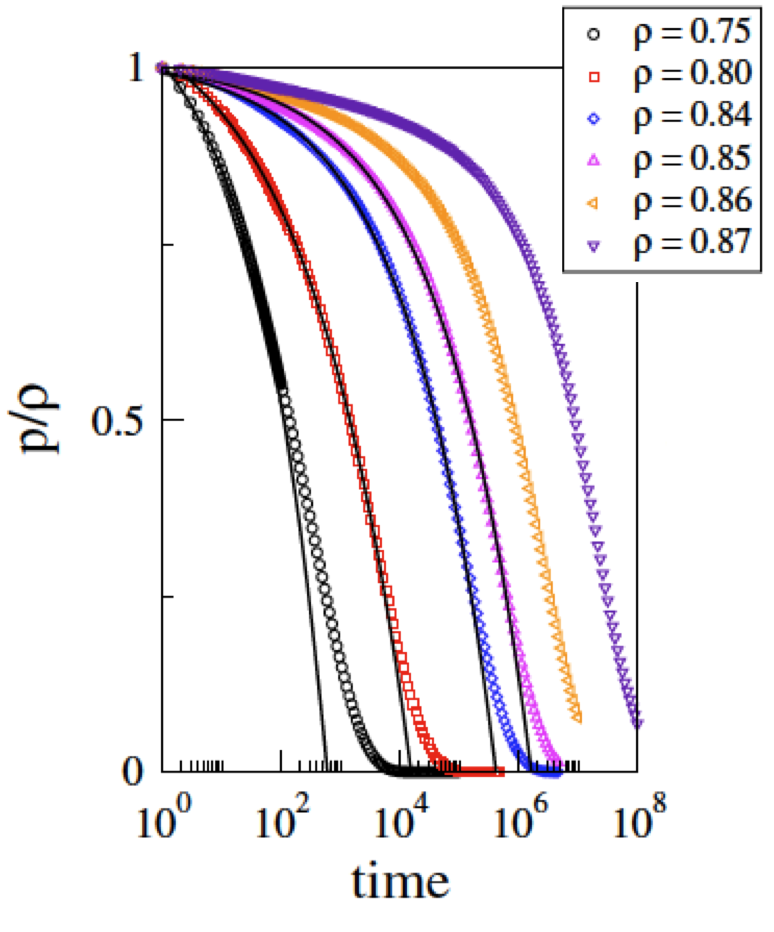}
\end{center}
\caption{\label{fig:review}
Normalized density of persistent particles $\<\rp\>/\rho$ 
for different values of the density, as indicated.
For $\rho \leq 0.85$, $\<\rp\>/\rho$ is well described by the von Schweidler law, $\<\rp\>/\rho = f_0- (t/\tau)^b$,
with $f_0 = 1$ and $b \simeq 0.3$. At short times, the dynamical susceptibility grows as $t^p$, with $p\simeq 0.61$. Adapted from Ref. \cite{PRL2011}.
}
\end{figure}
Accordingly, the normalized average value of such a correlation function, $\frac{\langle p(t)\rangle}{\rho}$, 
represents the dynamic order parameter for relaxation.
Fig. \ref{fig:review} shows that, in a large time window and for $\rho \leq 0.85$, 
$p(t)$ is well described by the von Schweidler law:

\begin{equation} \label{eq:Schweidler}
 \frac{\langle p(t) \rangle}{\rho}=f_{0}-(t/\tau)^b,
\end{equation}
with $b\simeq 0.3$ and $f_{0}\simeq 1$, whereas for larger time a stretched exponential
fit works better. 
Being the relaxation time $\tau$ defined as $\frac{\langle p(\tau) \rangle}{\rho}=1/e$, 
Fig. \ref{fig:xitau}b shows $\tau$ to diverge approaching the transition of structural arrest, 
$\tau(\rho) \propto (\rho_{ka}-\rho)^{-\lambda_{\tau}}$, with $\lambda_{\tau} \simeq 4.7$,
consistent with the result of Ref. \cite{FranzMulletParisi}.

\section{Dynamical Heterogeneities}
\label{DH}
Fig.~\ref{fig:plots} show that as less and less persistent particles survives, spatial correlations
between them emerge. These correlations are quantified by the dynamical susceptibility $\chi_4(t)$,
related to the fluctuations of $\rp$,
\begin{equation}
\label{eq:chip}
\chi_4(t)=\frac{V}\rho\left(\<\rp(t)^{2}\>-\<\rp(t)\>^{2}\right),
\end{equation}
and to the volume integral of the spatial correlation function between persistent particles at time $t$,

\begin{equation}
\label{eq:chi_gr} 
\chi_4(t) = \frac{1}{\rho V}\sum_{i,j}^{V}g_4(r,t)
\end{equation}
where

\begin{equation}
\label{eq:gr}
g_4(r,t)=\<n_{i}(t)n_{j}(t)\>-\<n_{i}(t)\>\<n_{j}(t)\>,\,\,\, r = |i-j|.
\end{equation}
The spatial decay of $g_4(r,t)$ defines the dynamical correlation length $\xi_{4}(t)$.
In the following we describe the behaviour of $g_4(r,t)$, $\xi_{4}(t)$ and  $\chi_4(t)$.

\subsection{Four-point correlation function}

\begin{figure}[t!]
\begin{center}
\includegraphics*[scale=0.45]{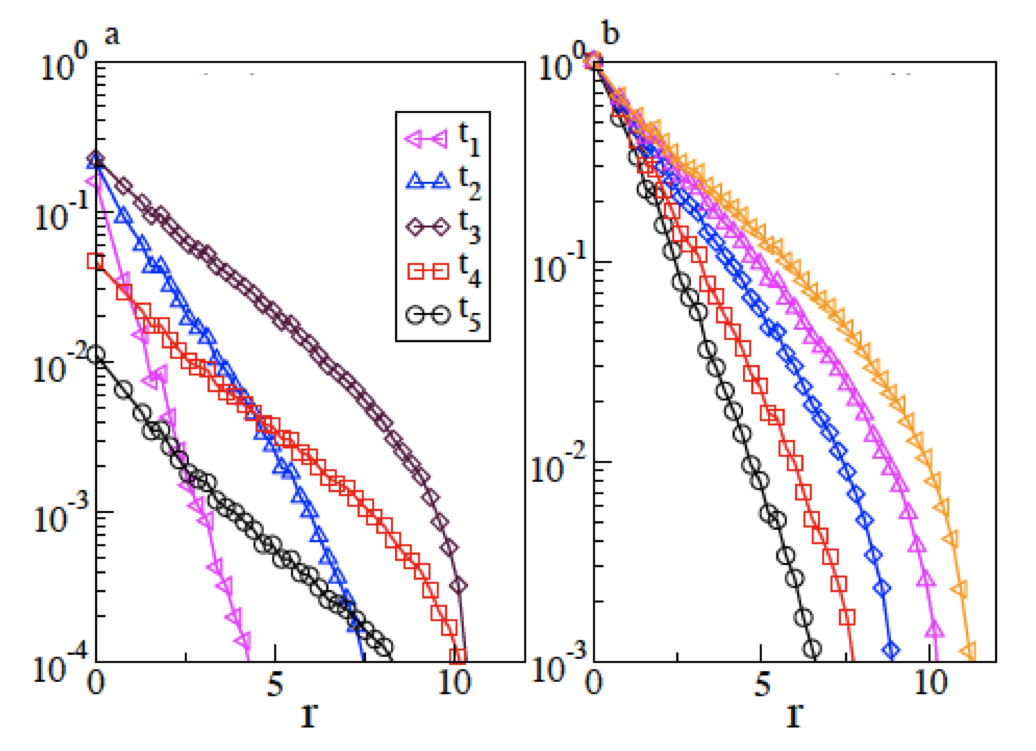}
\end{center}
\caption{\label{fig:g4_KA}
Panel a: Correlation function between persistent particles $g_{4}(r,t)$ as a function of the distance $r$ at $\rho=0.85$ and for different
times $t_1=10^2$, $t_2=5.3\cdot 10^3$, $t3=t^{*} _{\chi}=5.6 \cdot 10^5$, $t_4=2.8 \cdot 10^6$, $t_5=4.6 \cdot 10^6$. 
Panel b: $g4(r,t\simeq t^{*} _{\chi})$ as a function of the distance $r$ at different values of the density. From Ref. \cite{thesis}.
}
\end{figure}

In Fig. \ref{fig:g4_KA}a we show $g_4(r,t)$ at different times as a function of $r$. 
It clarifies that the spatial extension of the correlations 
grows until an intermediate time and then decreases at larger time. Its value at $r=0$,
 $g_4(0,t)=\langle p(t) \rangle (1-\langle p(t) \rangle)$, is determined by Eq. \ref{eq:gr}.

The behaviour of $g_4(r,t)$ is consistent with a scaling form such as:
 
 \begin {equation} \label {eq:gr_fit}
 g_4(r,t)=A(t) \frac{e^{r/\xi_{4}(t)}}{r^{d-2+\eta} } ,
 \end {equation}
 where $\xi_{4}(t)$ is the dynamical correlation length (whose behaviour
 will be discussed in the next session) and $A(t)$ is the amplitude.
 
 Fig. \ref{fig:g4_KA}b shows $g_4(r,t)$ normalized by its value at $r=0$ as a function of $r$, for different densities and at 
 intermediate time (roughly of the order of $t^{*}_{\chi}$). As expected,
 $g_4$ becomes increasingly long ranged when the transition of structural
 arrest is approached.

\subsection {Dynamical correlation length}
\begin{figure}[t!]
\begin{center}
\includegraphics*[scale=0.5]{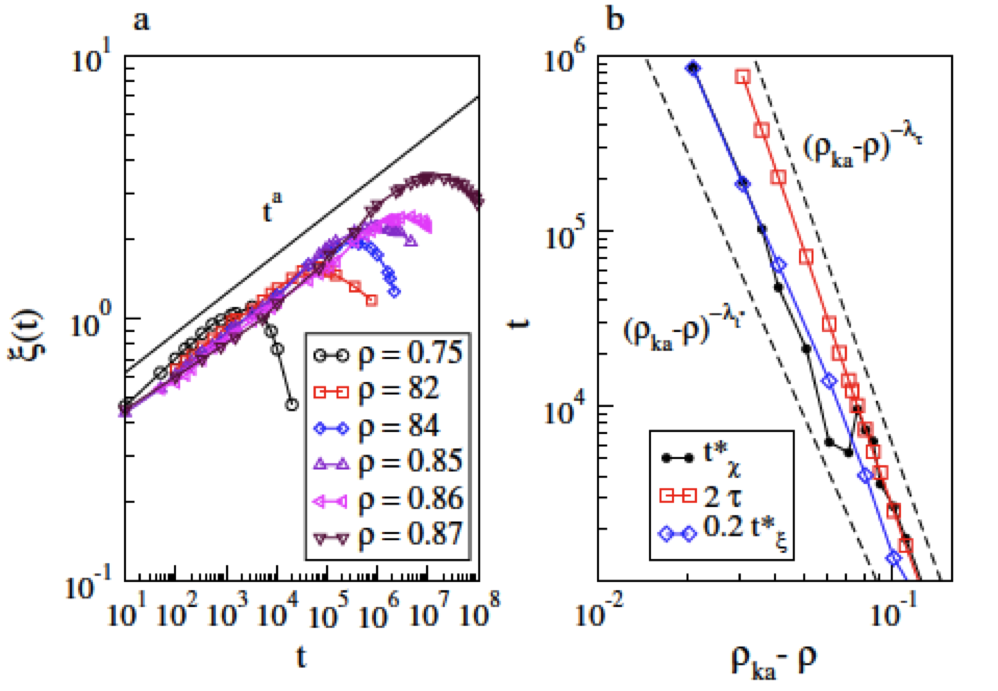}
\end{center}
\caption{\label{fig:xitau}
Panel a: dynamical correlation length for different values of the density.
Panel b: divergence of the relaxation time $\tau$, of the time where the correlation length acquires its maximum value $t^*_\xi$, and
of the time where the dynamical susceptibility acquires its maximum value, $t^*_\chi$. At low density, $t^*_\chi \propto \tau$,
while at high density $t^*_\chi \propto t^*_\xi$. Errors on $t^*_\xi$ and $t^*_\chi$ are of the order of $5\%$. From Ref. \cite{PRL2011}.
}
\end{figure}

 From the data relative to $g_4(r,t)$, we have extracted the correlation 
 length $\xi_{4}(t)$ via an exponential fit of the initial decay.
 The behaviour of $\xi_{4}(t)$  is illustrated in Fig.~\ref{fig:xitau}a for different values of the density,
 and is well described by
\begin{equation}
 \xi_4(t) \propto t^a \exp\left(-at/t^*_\xi\right).
\label{eq:xi}
\end{equation}

Accordingly, at short times $\xi(t)$ grows as $t^a$ with $a\simeq 0.156$ , 
and then it decreases after reaching its maximum value $\xi_4 ^*$ at time $t^*_\xi$.
We find that the time diverges 
as $t^*_\xi \propto \left(\rho_{ka}-\rho\right)^{-\lambda_{t^*_\xi}}$, 
with $\lambda_{t^*_\xi} = 3.8 \pm 0.1$.  

\begin{figure}[t!]
\begin{center}
\includegraphics*[scale=0.45]{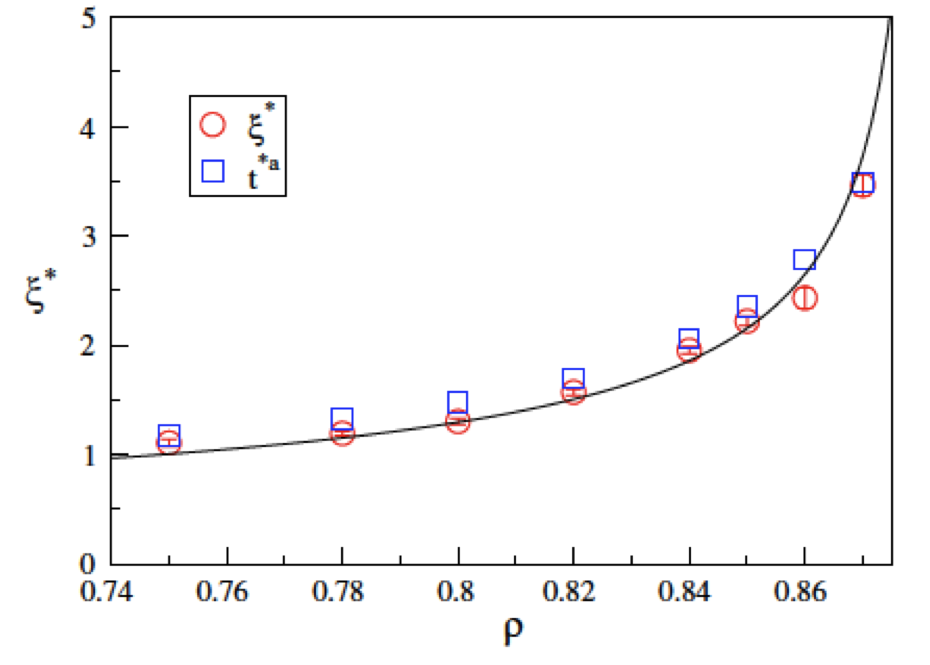}
\end{center}
\caption{\label{fig:lengths}
Dynamical correlation length at $t = t^*$, and prediction of the diffusing defect picture,
$\xi^* \propto t^{*a} \propto \tau^q$, $q = a \lambda_\tau/\lambda_{t^*_\xi}$. The full line is a $(\rka-\rho)^{-\nu}$, $\nu \simeq 0.54$
(we fix $\rho_{ka} = 0.881$ as estimated from the divergence of the relaxation time). From Ref. \cite{PRL2011}.
}
\end{figure}
\begin{figure}[t!]
\begin{center}
\includegraphics*[scale=0.45]{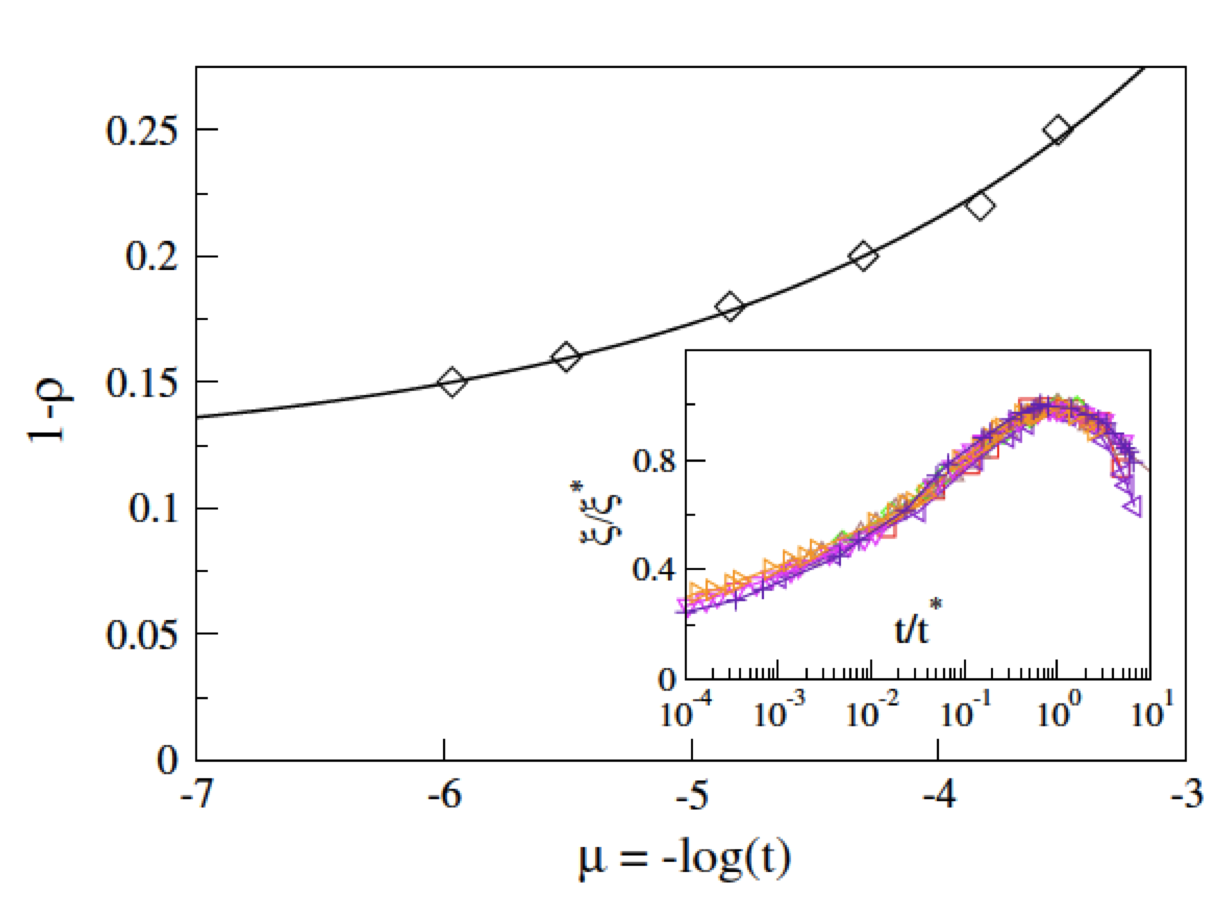}
\end{center}
\caption{\label{fig:widom}
Main panel: Widom line in the density, chemical potential plane. Circles indicates the time $t^*_\xi$
where the correlation length reaches its maximum value, at each value of the density. The continuous line corresponds
to $(1-\rka) \propto (t/t^*)^{a/\nu}$, and suggests that the system approaches a critical point at $\mu = -\infty$, and $\rho = \rka$.
Inset: scaling of the dynamical susceptibility for $7$ values of the density, in the range $0.78$--$0.87$. From Ref. \cite{PRL2011}.
}
\end{figure}

From this result and from Eq. \ref{eq:xi} it follows that the maximum dynamic correlation length diverges as
 $\xi_4^* \propto {t^*_\xi}^a \propto (\rho_{ka}-\rho)^{-\nu}$, with $\nu = a \lambda_{t^*_\xi} \simeq 0.54$, in very good agreement
 with data (Fig.~\ref{fig:lengths}). Note here a very important 
 result: contrary to what is expected, we find that  $\tau$ and $t^{*} _{\xi} $
are not proportional but seems to diverge with different exponents, $\lambda_{\tau} > \lambda_{t^{*} _{\xi}}$,
approaching the transition of structural arrest.
The presence of a growing correlation length suggests that the system is approaching a critical point as the density increases.
This scenario is conveniently described interpreting $\mu = -\log(t)$ as a chemical potential for the persistent particles,
considering that the density of persistent particles 
monotonically decreases as time advances. The line where
the correlation length reaches its maximum value in the $\mu$--$\rho$ plane
can therefore be interpreted as a Widom line, which in a second order transition ends at the critical point.
The results of Fig.~\ref{fig:widom} suggest the presence of a critical point located at $\rho = \rka$ and $\mu = -\infty$,
where the correlation length diverges. Actually, the Widom line will possibly  bend,
ending at $\rho = 1$ where the transition is known to occur in the thermodynamic limit.
Such an approach may open the way to a renormalization group treatment of the glass transition.

 \subsection{Dynamical Susceptibility}

\begin{figure}[t!]
\begin{center}
\includegraphics*[scale=0.4]{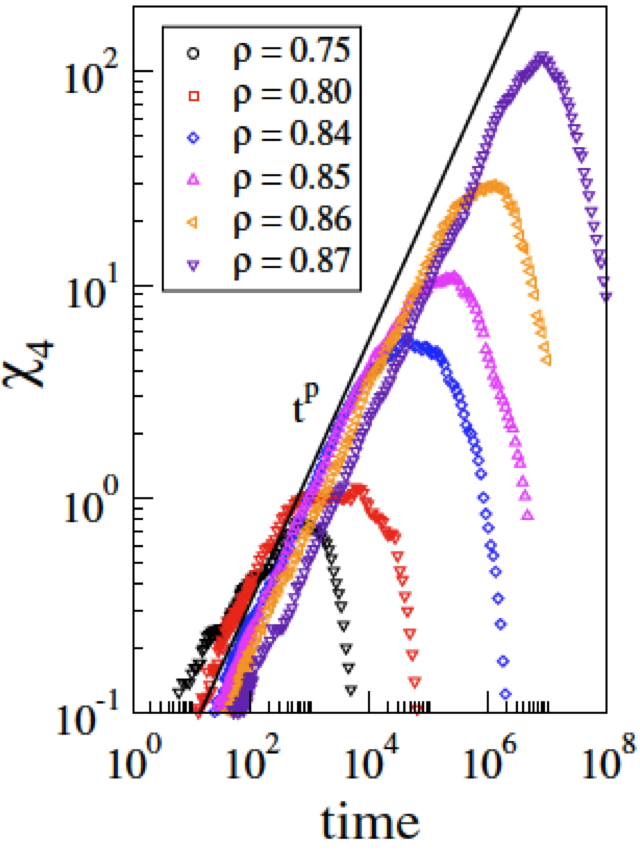}
\end{center}
\caption{\label{fig:chi4}
Dynamical Susceptibility, $\chi_4(t)$, for different values of the density,
as indicated.  At short times, $\chi_4(t)$ grows as $t^p$, with $p\simeq 0.61$. Adapted from Ref. \cite{PRL2011}.
}
\end{figure}

The emergence of an increasingly
heterogeneous dynamics is clearly signaled by the
dynamical susceptibility $\chi_4(t)$, shown in Fig. \ref{fig:chi4} for different values of the density.
Qualitatively the behaviour of $\chi_4(t)$ appears similar to the one observed for $\xi_{4}(t)$,
initially growing as $\chi_4(t) \propto t^p$, with $p\simeq 0.6$
and then decreasing  after reaching its maximum value $\chi_4^*$ at a time $t^*_\chi$.
The decoupling between $t^*_\xi$ and $\tau$ strongly influences $\chi_4^*(\rho)$ and the time $t^*_\chi(\rho)$,
leading to a complex behaviour.
In facts, inspired by Eq. \ref{eq:chi_gr} and Eq. \ref{eq:gr_fit}
we find that $\chi_4(t)$ is well approximated by

\begin{equation} \label {eq:chi_fit} 
\chi_4(t) \propto A(t) \xi(t)^{2-\eta} = \left[\langle p(t) \rangle (1-\langle p(t) \rangle)\right] \xi(t)^{2-\eta},
\end{equation}
where we have estimated $A(t)=g_4(0,t)$ and
$\eta \simeq 0$ consistent with Ref. \cite{etazero}.
Accordingly, the behaviour of the susceptibility is essentially given by the product of two competitive factors,
the amplitude and the correlation length.
Indeed, at low densities $t^*_\xi \gg \tau$, the amplitude dominates
and Eq.~\ref{eq:chi_fit} predicts $t^*_\chi \propto \tau$.
By contrast, at high density $t^*_\xi \ll \tau$, and
the maximum of the susceptibility $\chi_4^*$ occurs at $t^*_\chi \propto t^*_\xi$.
Such behavior of $t^*_\chi$ is apparent in Fig.~\ref{fig:xitau}: at intermediate density
$t^*_\chi$ shows a clear crossover separating the two asymptotic regimes, where $t^*_\chi$
scales as $\tau$ in the limit of low-density and as $t^*_\xi$ in the limit of high density.
Asimptotically, when $t^*_\chi \propto \tau$, the maximum of the susceptibility
scales as $\chi_4^* \propto \tau^{2a} \propto (\rho_{ka}-\rho)^{-\gamma}$,
with $\gamma = 2a\lambda_\tau$, in agreement with our results.
Conversely, when $t^*_\chi \propto t^{*}_{\xi}$, 
 we have $\chi_4^*\propto (t^*_{\xi})^{2a} \propto (\xi_4^{*})^{2} \propto (\rho_{ka}-\rho)^{-q}$, with
$q =2a\lambda_{t^*_\xi}$.

\section{Diffusing defects}
\label {dd}
The results described so far are rationalized in the
diffusing defects paradigm~\cite{ToninelliE,reviewBB,Shlesinger, ChandlerPRX}, where
the relaxation is ascribed to the presence of possibly extended diffusing defects, with density $\rho_d$.
The diffusing defects picture provides that
at short times
the correlation length grow as $\xi_4(t) \propto t^{1/x}$. 
The number of distinct sites visited by a defect increases as 
$n_v(t) \propto t^{d_f/x}$, where
$d_f$ and $1/x$ are respectively the fractal dimension and the diffusion exponent
characterizing the defect walk.
Thus, the total number of distinct sites visited by the defects at time $t$
is proportional to $\rho_d n_v(t)$.
These assumptions are expected to hold at least at short times, i.e.  before defects interact.  
Under this condition, each site visited by a defect corresponds
to a particle which first moves from its original position.
Thus, we expect that at short times the total number of distinct visited sites scales as the density of
particles which have already relaxed at that time (the non-persitent particles), $1- \frac{\langle p(t) \rangle} {\rho}$.
From the decay observed for the density of persistent particles,  we can infer
$1- \langle p(t) \rangle/\rho = (1/\tau)^b t^b \propto  \rho_d n_v(t) \propto \rho_d t^{d_f/x}$.
Therefore this picture
reproduces the von Schweidler law, and relates the density of defects $\rho_d$ with the relaxation time
\cite{nota2}, $\rho_d \propto \tau^{-b}$.
By comparing the correlation length scaling provided by diffusing defect
picture with the observed short time behaviour,
$\xi_4(t) \propto t^a$ (Fig. \ref{fig:lengths}), we can conclude $1/x=a$ and $d_f=b/a\simeq2$.
Then we find that defects have  a sub--diffusive nature, 
although they conserve the same fractal dimension of usual random walkers.
We suggests this may be ascribed to defects
that behave as random walkers, although characterized by a fat--tail waiting time distribution and,
possibly, by weak spatial correlations which slows down the diffusion. 
In fact, it has been proved that these factors do not affect the fractal
dimensions of the walkers ~\cite{Yuste, attractiveRW}.
Dealing with the susceptibility, the diffusing defect picture predicts
 that at short times,
$\chi_4(t) \propto \rho_d n_v(t)^2 \propto t^{2b}$, which compared with our result
 allows to correctly estimate $p = 2b$.

\section{ Reverse percolation}
\label{percolation}
Looking for a geometrical interpretation of the relaxation process, 
we were inspired by the analogy with chemical gels. 
The mechanical rigidity of chemical gels, in fact, arises from a percolating network 
of polymers which acts as a backbone in a liquid media. 
Since the bond of such network are of a chemical nature, 
relaxation cannot spontaneously occur and
 the system is solid at any time-scale.
 However relaxation may be induced by external factors,
 such as diffusing enzymes able to cut the bonds they meet \cite{Abete,Sidoravicius}:
 in this case, the system stays
solid until the network survives, 
but it relaxes and becomes liquid at larger time-scales, when the network disappears. 
Similar features have been also found for the stress-bearing network of contact forces in jammed granular materials \cite{PM, PRE, GM}. 
In the present case the glass former may be though as rigid on time-scales smaller than the relaxation time, $\tau$. 
We suppose that during this time a percolating cluster of persistent particles plays the role of the physical backbone in gels;
here the bonds are of a dynamical kind, i.e. we consider that two particles $i$ and $j$
 are bonded in the interval $[0,t]$ if they are nearest neighbours and persistent in this time interval.
Defects, instead, play the role of enzymes and progressively destroy the cluster. 
A reverse dynamical percolation transition is expected for time-scales of the order of the relaxation time. 
In fact, as the absence of the percolating cluster leads to the loss of rigidity, 
we expect that this transition is related to the relaxation process.

\begin{figure}[t!]
\begin{center}
\includegraphics*[scale=0.55]{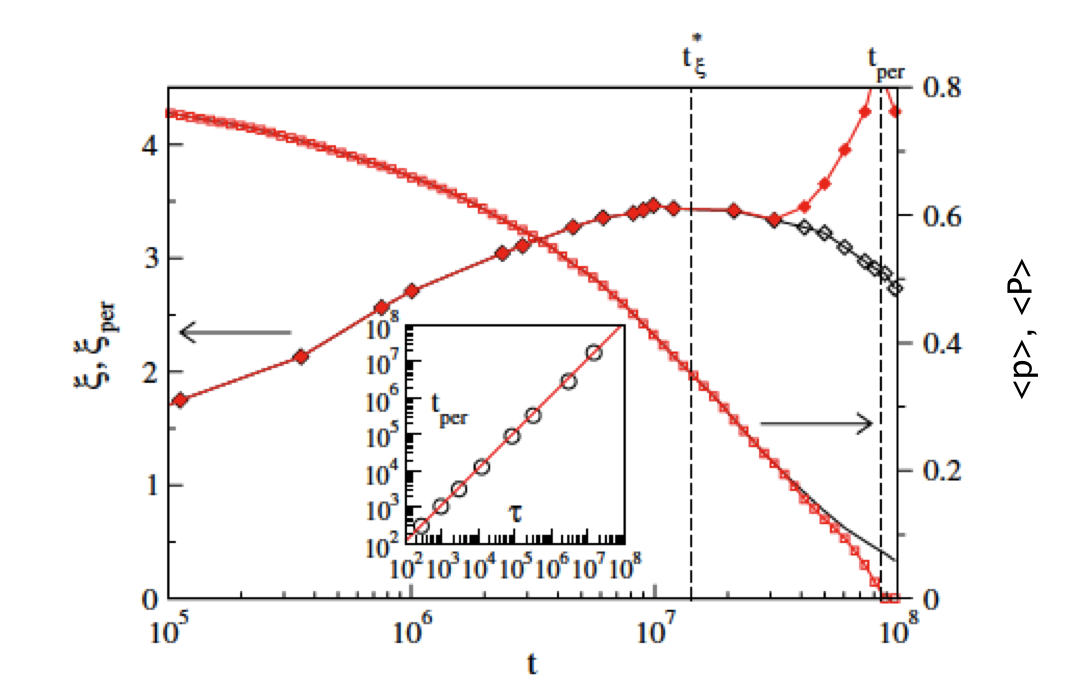}
\end{center}
\caption{\label{fig:rho85}
Percolation transition at $\rho = 0.87$. Left axis:
dynamical correlation length $\xi$ (empty diamonds) and percolation correlation
length $\xip$ (full diamonds). Right axis: density of persistent particles $\<\rp\>$ (full line)
and strength of the percolating cluster $\<P\>$ (squares).
The vertical dashed lines mark $t^*_\xi$ and $\tp$, which is proportional to $\tau$ (inset). From Ref. \cite{PRL2011}.
}
\end{figure}

Data shown in Fig. \ref{fig:rho85}  confirm our hipotesys. For density, beyond the onset of glassy dinamics,
a cluster of persistent particles always spans the system at short times.
We indicate with $P(t)$ its strength, i.e. the density of persistent particles belonging to such cluster. 
$\langle P(t) \rangle$ vanishes at a time $t_{per}$ which is 
found to scale with the relaxation times, $\tau$, as the density increases (Fig. \ref{fig:rho85}, inset). 

The figure also reveals that the cluster strength overlaps with the total density
of persistent particles, $\langle P(t) \rangle \simeq  \langle p(t) \rangle$, up to large times.
This means that in this interval the percolating cluster is the only cluster present.
At larger time $p(t)$ slowly decays, while $P(t)$ vanishes. 
This is due to the onset of finite clusters with a broad size distribution, that give contributions to $p(t)$, but not to $P(t)$.
This circumstance may explain the crossover  observed for the dynamic correlation function (see Fig. \ref {fig:review}): 
at short time, the decay of $\langle p(t) \rangle/\rho$  is characterized by a single relaxation time,
which is the life-time of the percolating cluster, and this leads to the von Schweidler law (Eq. \ref{eq:Schweidler}).
By contrast, the broad spectra of finite cluster life-time determine the stretched exponential decay at large times.

To better understand the geometrical properties of this process,
we investigate the correlation length, $\xi_{per}(t)$, which 
is defined by the percolative correlation function $g_{per}(r,t)$:
 
 \begin{equation} \label{eq:g_per}
 g_{per}(r,t)=g_{pc}(r,t)-\langle P(t) \rangle^2= \langle n_i(t)n_j(t) \rangle -\langle P(t) \rangle^2 
 \end{equation}
 where the \textit{pair-connected} correlation function $g_{pc}(r,t)$ is limited
  to the pairs of particles that belong to the same dynamical cluster. $g_{pc}(r,t)$ can be also expressed as
  $g_{pc}(r,t)=P_{i,j}^{f}(r,t)+P_{i,j}^{\infty}(r,t)$ where $P_{i,j}^{f}(r,t)$ and $P_{i,j}^{\infty}(r,t)$
 are the probabilities that two sites $i$ and $j$ belong to the same finite cluster, or to the percolating cluster respectively.
Accordingly, if the percolating cluster is absent  $g_{per}(r,t)=P_{i,j}^{f}(r,t)$ and $\xi_{per}(t)$ 
measures the typical size of finite clusters. Conversely, if finite clusters are negligible, then 

\begin{equation} \label{g_per_inf}
g_{per}(r,t)=P_{i,j}^{\infty}(r,t)-\langle P(t) \rangle^2 
\end{equation}
measures the extension of the density fluctuations within the percolating cluster. 
In our case, at short time, $\langle P(t) \rangle \simeq  \langle p(t) \rangle$ and $P_{i,j}^{\infty}(r,t)\simeq \langle n_i(t)n_j(t) \rangle_{|i-j|=r}$,
because almost all persistent particles belong to the percolating cluster, making the connectedness condition negligible \cite{Coniglio}.
Inserting these equalities in Eq. \ref{g_per_inf} and comparing it with the definiton of the four point correlation function, $g_4(r,t)$ (see Eq. \ref{eq:gr}),
we find that $g_{per}(r,t)\simeq g_{4}(r,t)$, and consequently $\xi_{per}(t)\simeq\xi_4(t)$.
Indeed, Fig.~\ref{fig:rho85} confirms that the dynamical correlation length coincides with the percolative length,
 as long as finite clusters are negligible.
The percolative length is affected by the two timescales characterizing
the glassy dynamics, the time $t = t^*_\xi$ where the dynamical length reaches its maximum value, and the percolating time related
to the relaxation of the system. At high densities, this makes $\xip$ non monotonic, as in Fig.~\ref{fig:rho85}.

\section{Conclusion}
We have shown that in the KA
model relaxation process and DHs
are characterized by two different timescales $\tau$ and $t^*_\xi$,
which implies that they are 
less tangled than expected.
We explain this feature in the diffusing defect picture, 
where we relate the relaxation process 
to a reverse percolation transition, 
obtaining a geometrical interpretation of  the different time-scales.
Accordingly, to their definitions, $\tau$ occurs when a given
large fraction of all sites has relaxed, while $t^*_\xi$ occurs
when the correlations between the persistent particles decreases.
Indeed, nothing forbids the correlations 
to decrease before a large fraction of all sites has relaxed, $t^*_\xi<\tau$ .
However, in models of defects behaving as random walkers, one finds that 
 $\tau \propto t^*_\xi$ ~\cite{ToninelliE}.
Therefore, one may speculate that, in our case, the decoupling  between the two time-scales
is due to a complex nature of defects. 
 For instance, they may be non-conservative and characterized by birth and death rate 
with a constant average number, or they may not diffuse as
random walkers.


\begin{thebibliography}{30}
\bibitem{nota}
$n_{i}(t)$ does not explicitly shows  the two-time structure typical of dynamic correlation functions.
By contrast, it is even more than a two-time function, since it keeps trace of the whole
history in the interval $[0,t]$: indeed, it can be written as multi-point function, $n_{i}(t)=\Pi_{t'=0} ^{t'=t} m_{i}(t)$, where
$m_{i}(t)$ is the usual occupation number of site $i$ at time $t$
($m_{i}(t)=1(0)$ if site $i$ is (is not) occupied by any particle at time $t$).
\bibitem{nota2}
We are assuming that the diffusion coefficient of defects $D_d(\rho)$ varies slowly in the investigated range of density.
\end{thebibliography}
\end{document}